\documentclass{article}
\usepackage{spconf,amsmath,graphicx}

\usepackage{url,times,booktabs, tabularx}
\usepackage{multirow}
\usepackage{caption}
\usepackage{subcaption}
\usepackage{enumitem}

\title{A SEQUENCE MATCHING NETWORK FOR POLYPHONIC SOUND EVENT LOCALIZATION AND DETECTION}
\name{Thi Ngoc Tho Nguyen$^{\star}$ \qquad Douglas L. Jones$^{\dagger}$  \qquad Woon-Seng Gan$^{\star}$ \sthanks{This material is based on research work supported by the IAF-ICP: Singtel Cognitive and Artificial Intelligence Lab for Enterprises@NTU under the Research Theme on Edge Intelligence.}}

\address{$^{\star}$ Department of Electrical and Electronic Engineering,\\
    Nanyang Technological University, 639798, Singapore\\
    $^{\dagger}$ Department of Electrical and Computer Engineering,\\
    University of Illinois at Urbana-Champaign, IL 61801, United States }   

\begin{document}
\ninept
\maketitle
\begin{abstract}
Polyphonic sound event detection and direction-of-arrival estimation 
require different input features from audio signals. 
While sound event detection mainly relies on time-frequency patterns, direction-of-arrival estimation relies on magnitude or phase differences between microphones. 
Previous approaches use the same input features for sound event detection and direction-of-arrival estimation, and train the two tasks jointly or in a two-stage transfer-learning manner. We propose a two-step approach that decouples the learning of the sound event detection and directional-of-arrival estimation systems. In the first step, we detect the sound events and estimate the directions-of-arrival separately to optimize the performance of each system. In the second step, we train a deep neural network to match the two output sequences of the event detector and the direction-of-arrival estimator. This modular and hierarchical approach allows the flexibility in the system design
, and increase the performance of the whole sound event localization and detection system. 
The experimental results using the DCASE 2019 sound event localization and detection dataset show an improved performance compared to the previous state-of-the-art solutions.  
\end{abstract}
\section{Introduction}
\label{sec:intro}

Polyphonic sound event localization and detection (SELD) has many applications in urban sound sensing~\cite{Salamon2017Cnn}, wild life monitoring~\cite{Stowell2016bird}, surveillance~\cite{Foggia2016Surveillance}, autonomous driving~\cite{nandwana2016car}, and robotics~\cite{valin2004localization}. The SELD task recognizes the sound class, and estimates the direction-of-arrival (DOA), the onset, and offset of a detected sound event~\cite{Adavanne2019seld}. Polyphonic SELD refers to cases where there are multiple sound events overlapping in time. 

SELD consists of two subtasks, which are sound event detection (SED) and direction-of-arrival estimation (DOAE). In the past decade, deep learning has achieved great success in classifying, tagging, and detecting sound events~\cite{virtanen2018computational}. The state-of-the-art SED models are often built from convolutional neural networks (CNN)~\cite{Salamon2017Cnn}, recurrent neural networks (RNN)~\cite{parascandolo2016recurrent}, and convolutional recurrent neural networks (CRNN)~\cite{Adavanne2019seld, cakir2017convolutional}. DOAE tasks for small-aperture microphone arrays are often solved using signal processing algorithms such as minimum variance distortionless response (MVDR) beamformer~\cite{Capon1969mvdr}, multiple signal classification (MUSIC)~\cite{schmidt1986multiple}, and steered-response power phase transform (SRP-PHAT)~\cite{salvati2014incoherent}. To tackle the multi-source cases, many researches exploit the non-stationarity and sparseness of the audio signals to find the single-source time-frequency (TF) regions on the spectrogram to reliably estimate DOAs~\cite{mohan2008localization, tho2014robust, Griffin2012doa}. Recently, deep learning has also been successfully applied to DOAE tasks~\cite{xiao2015learning, adavanne2018doaRcnn}, and the learning-based DOA models show good generalization to different noise and reverberation levels. However, the angular estimation error is still high for multi-source cases.

An SELD system is expected to not only able to perform its two subtasks well but also able to correctly associate the detected sound events with the estimated DOAs. Hirvonen formulated the SELD task as multi-class classification where the number of output classes is equal to the number of DOAs times the number of sound classes~\cite{hirvonen2015classification}. Clearly, this approach is not scalable to a large number of DOAs and sound classes. Adavanne \emph{et al.} proposed a single-input multiple-output CRNN model called SELDnet that jointly detects sound events and estimates DOAs~\cite{Adavanne2019seld}. The model's loss function is a weighted sum of the individual SED and DOAE loss functions. The joint estimation affects the performance of both the SED and DOAE tasks. To mitigate this problem, Cao \emph{et al.} proposed a two-stage strategy for training SELD models~\cite{cao2019polyphonic}. First, a SED model using a CRNN architecture is trained by minimizing the SED loss function using all the available data. After that, the CNN weights of the SED model is transferred to the DOA model, which has the same architecture as the SED model. The DOA model is trained by minimizing the DOA loss function using only the data that have active sources. The SED outputs are used as masks to select the corresponding DOA outputs. This training scheme significantly improves the performance of the SELD system. However, the DOA model is still dependent on the SED model for detecting the active signals, and the network learns to associate specific sources with specific directions in the training data. In the DCASE 2019 challenge, the top solution trained four separated models for sound activity detection (SAD), SED, single-source and two-source DOAE respectively~\cite{Kapka2019seld}. This solution heavily used heuristic rules to determine the single-source and two-source segments of the signal 
to infer the sound classes and DOAs. This approach is not scalable when there are more than two overlapping sounds. 

We propose a novel two-step approach that decouples the learning of the SED and DOAE systems. In the first step, we use Cao's CRNN model~\cite{cao2019polyphonic} to detect the sound events, and a single-source histogram method~\cite{tho2014robust} to estimate the DOAs. In the second step, we train a CRNN model to match the two output sequences of the event detector and DOA estimator. The motivation of this approach is that overlapping sounds often have different onsets and offsets. By matching the onsets, the offsets, and the active segments in the output sequences of the sound event detector and the DOA estimator, we can associate the estimated DOAs with the corresponding sound classes. This modular and hierarchical approach significantly improves the performance of the SELD task across all the evaluation metrics. 

In addition, we propose a new output format for the SELD system that can handle the case where two sound events have the same sound class but different DOAs. This case cannot be resolved using the current output format of the state-of-the-art SELD systems~\cite{Adavanne2019seld, cao2019polyphonic}, which only allows one DOA estimate per sound class. 
We propose to estimate the sound events in each data frame directly. The frame-level sound event is defined as sound produced by a unique source and has two attributes: a sound class and a DOA. 
Each data frame can have zero, one, or multiple sound events. The network estimates the number of sound events together with their respective sound classes and DOAs. We introduce two new evaluation metrics that quantify the degree to which the DOAs are matched to the correct sound classes. We use the first-order ambisonic (FOA) format of the DCASE 2019 SELD dataset~\cite{Adavanne2019_DCASE} to evaluate our approach. The rest of our paper is organized as follows. Section II describes our proposed two-step SELD system. Section III presents the experimental results and discussions. 

\section{A two-step SELD system}
\label{sec:two_step}

\begin{figure*}[t]
\centering
\includegraphics[width=1\textwidth]{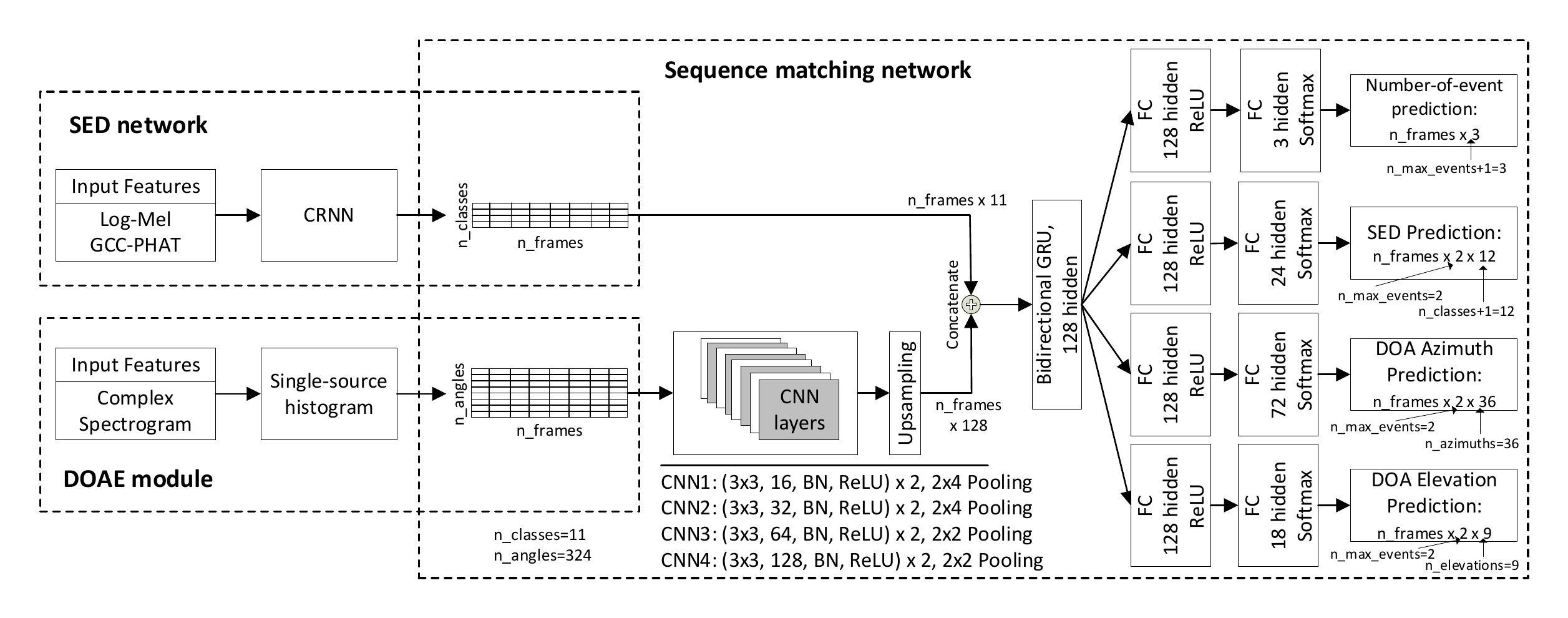}
\caption{Block diagram of the two-step sound event localization and detection. Step 1: SED network and DOA module generate SED and DOA output sequences. Step 2: Sequence matching network matches the sound classes and DOAs for detected sound events.$n_{frames}$ is the number of time frames of one training samples, $n_{classes}$ is the number of sound classes, $n_{angles}$ is the number of DOAs.}
\label{fig1:seld_block_diagram}
\end{figure*}

Figure~\ref{fig1:seld_block_diagram} shows the block diagram of the proposed two-step SELD system. The SED network is similar to the one proposed by Cao \emph{et al}~\cite{cao2019polyphonic}. The DOAE module uses a non-learning signal processing approach to robustly estimate the DOAs of sound sources regardless of the sound classes~\cite{tho2014robust}. The output sequences of the SED network and DOAE module are the inputs of the sequence matching network. The sequence matching network (SMN) uses CNN layers to reduce the dimension of the DOA inputs before concatenating them with the SED inputs. A bidirectional gated recurrent unit (GRU) is used to match the DOA and SED sequences. And fully connected (FC) layers are used to produce the final SELD estimates.

\subsection{Sound event detection}

For the SED task, we used the SED network proposed by Cao \emph{et al.}~\cite{cao2019polyphonic}. The SED network uses log-mel spectrogram and generalized cross-correlation phase transform (GCC-PHAT) as input features. GCC-PHAT is shown to improve the performance of the SED task when used in conjunction with log-mel spectrogram~\cite{cao2019polyphonic, adavanne2017sedspatial}. The audio signal is divided into windowed overlapping audio frames. Log-mel spectrum is computed for each audio frame for all the microphones. GCC-PHAT is computed for each audio frame for all the microphone pairs. 
The number of mel filters is chosen to match the size of the GCC-PHAT spectrum so that the log-mel and the GCC-PHAT features can be stacked along the microphone and microphone-pair dimension respectively. 
The original paper~\cite{cao2019polyphonic} does not use data augmentation. To further improve the SED performance, we use random cut-out augmentation for the input data~\cite{zhong2017random}. Random cut-out augmentation randomly masks a rectangular block spanning several frequency bands and time-steps of the input features. We use the same cut-out mask for all the channels. 

The SED network consists of $8$ CNN layers, $1$ bidirectional GRU layer, and $1$ FC layer. The SED is formulated as multi-label multi-class classification. Adam optimizer is used to train the network. We use the raw probability output of the SED network as the input to the sequence matching network in step $2$.  

\subsection{Direction-of-arrival estimation}

We use a single-source histogram algorithm proposed in ~\cite{tho2014robust} to estimate DOAs. The single-source histogram finds all the time-frequency (TF) bins that contains energy from mostly one source. A TF bin is considered to be a single-source TF bin when it passes all three tests: magnitude, onset, and coherence test. Magnitude test finds the TF bins that are above a noise floor to mitigate the effect of background noise. Onset test finds the TF bins that belong to direct-path signals to reduce the effect of reverberation in the DOA estimation. Coherence test finds the TF bins of which the covariance matrices are approximately rank-$1$. After all the single-source TF bins are found, the DOA at each bin is computed using the theoretical steering vector of the microphone array~\cite{tho2014robust}. These DOAs are discretized using the required resolution of azimuth and elevation angles. Subsequently, these DOAs are populated into a histogram, which is smoothed to reduce the estimation errors. The final DOA estimates are the peaks of this histogram. 

\begin{figure}[tb]
\centering
\includegraphics[width=0.5\textwidth]{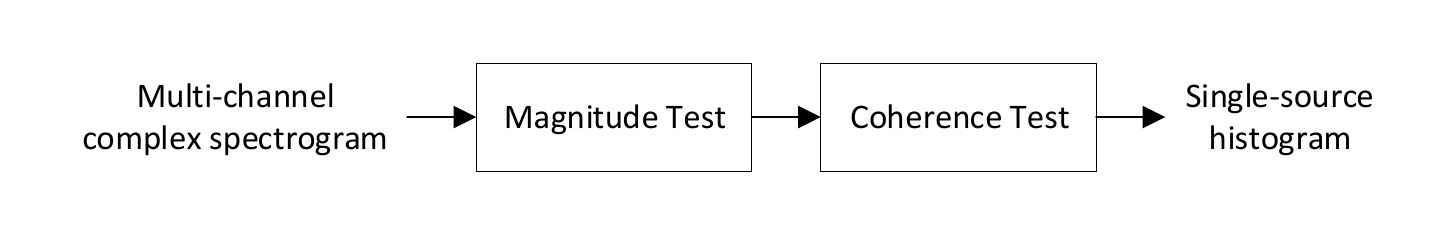}
\caption{Block diagram of single-source histogram algorithm.}
\label{fig2:histogram}
\end{figure}

We compute one histogram for each time frame. Since the SELD dataset has maximum two overlapping sources and moderate levels of reverberation, the onset test is not used. In addition, we do not use any smoothing function on the histogram. The block diagram of the single-source histogram algorithm is shown in Fig.~\ref{fig2:histogram}. The $2D$ histogram is vectorized into a single vector for each data frame. 

\subsection{Sequence matching network}

\begin{figure}[tb]
\centering
\includegraphics[width=0.45\textwidth]{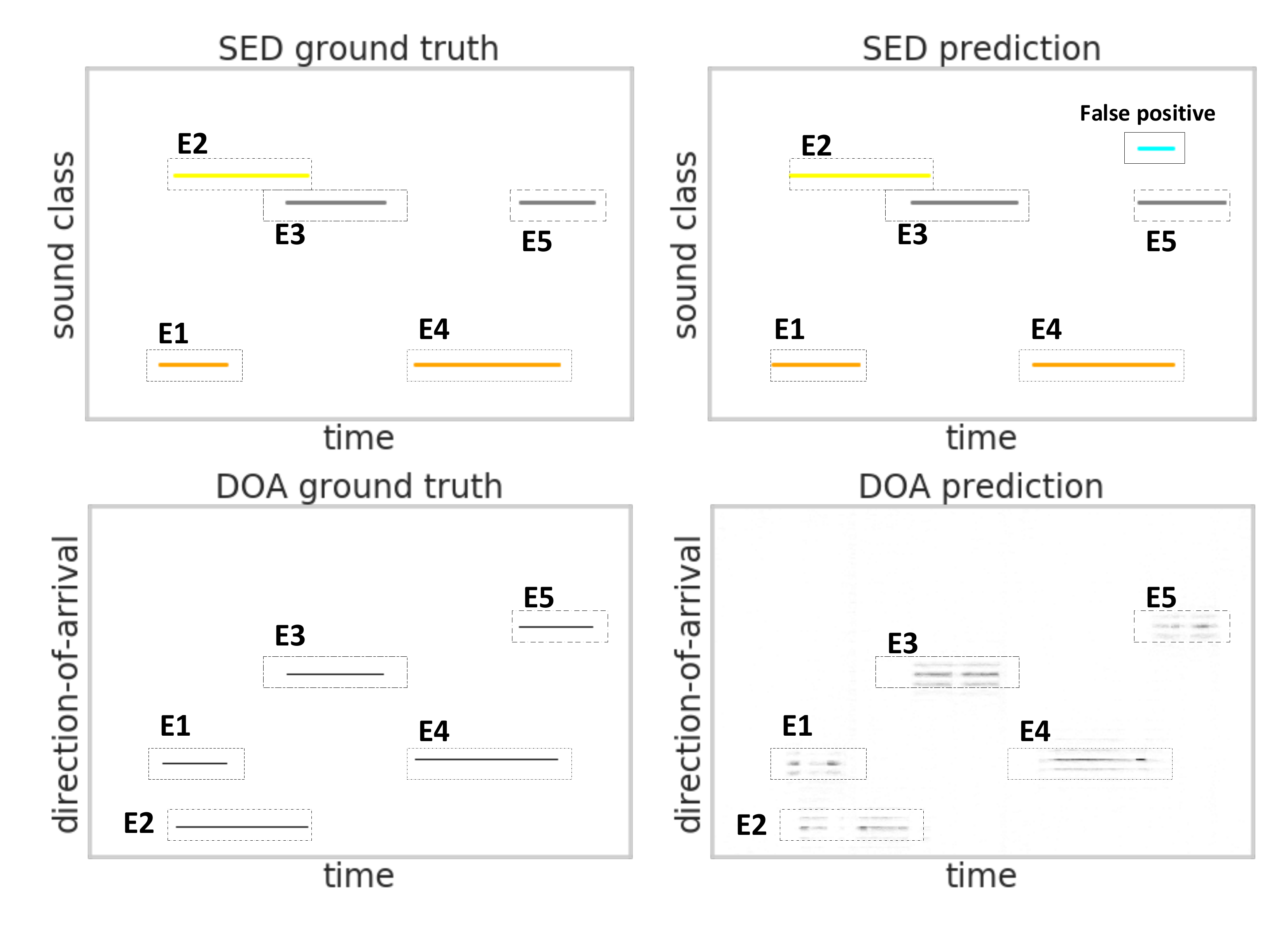}
\caption{Ground truth and prediction sequences of SED and DOAE.}
\label{fig3:matching}
\end{figure}

Figure~\ref{fig3:matching} shows the output and the ground truth sequences of the SED network and the DOAE module for a $5$-second audio segment taken from the DCASE dataset. The outputs of the SED network show the detected sound events with their corresponding sound classes. The outputs of the DOAE module show the detected sound events with their corresponding DOAs. We binarize the SED output using a threshold of $0.3$ for plotting purposes. Fig.~\ref{fig3:matching} shows that it is possible to visually match the onsets, offsets, and active segments of the sound events in the two output sequences in order to associate the sound classes, DOAs to the detected events. The sequence matching network (SMN) is trained to do this task automatically. 

The details of the SMN is shown in Fig.~\ref{fig1:seld_block_diagram}. Because the number of the discretized DOAs is often much larger than the number of sound classes, we use several CNN layers to reduce the dimension of the DOA input features before concatenating them with the SED input features. A bidirectional GRU layer is used to perform sequence matching. FC layers are used to produce desired outputs. 

\begin{figure}[tb]
\centering
\includegraphics[width=0.45\textwidth]{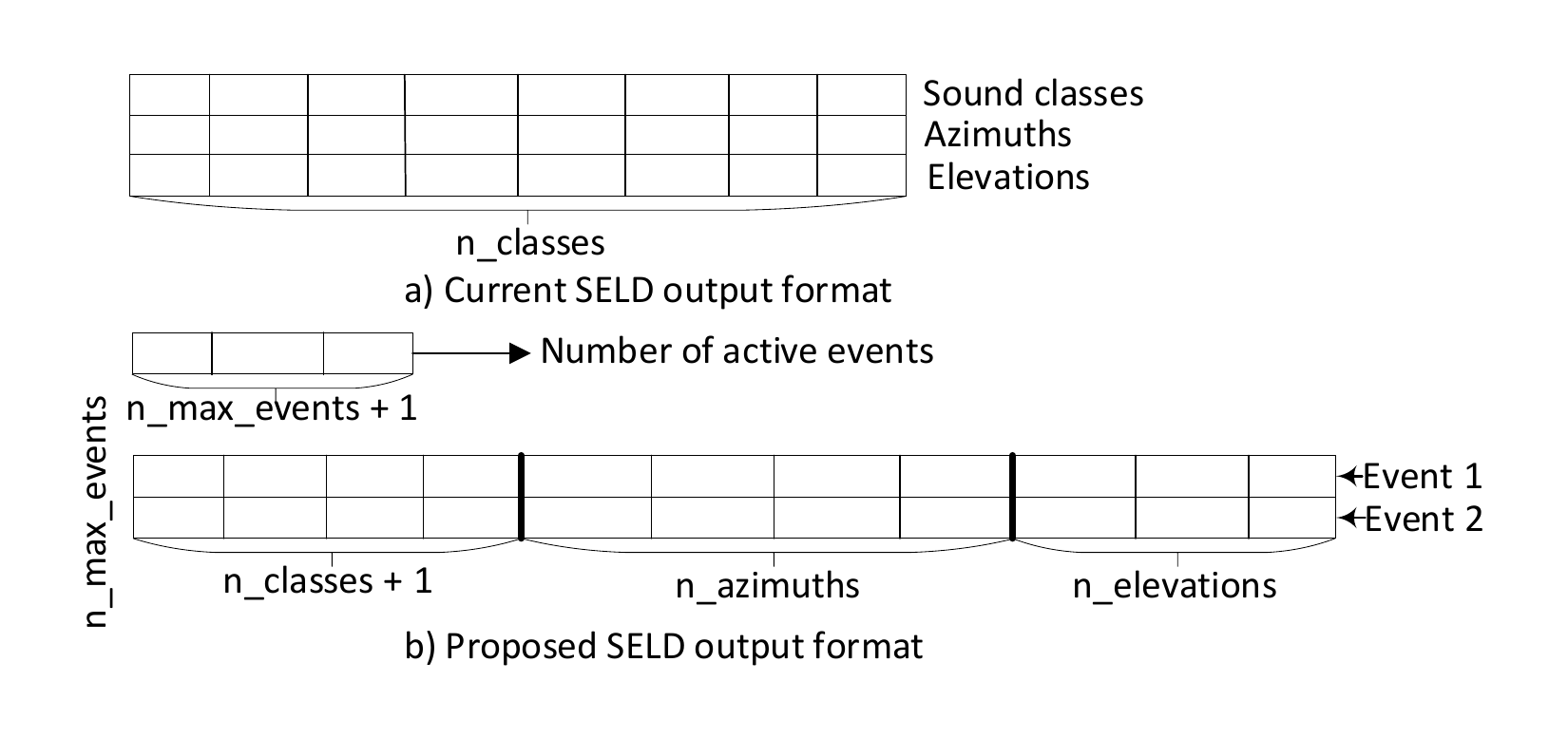}
\caption{Output formats for SELD task.}
\label{fig4:output_format}
\end{figure}

The SELDnet~\cite{Adavanne2019seld} and the two-stage SELD system~\cite{cao2019polyphonic} use the multi-label multi-class classification for SED and regression for DOAE as shown in Fig.~\ref{fig4:output_format}a. This output format could not handle the case where two overlapping sound events have the same sound class but different DOAs. This situation is common in indoor environment for \emph{speech} class. We propose a new SELD output format as shown in Fig.~\ref{fig4:output_format}b. For each time frame, we do multi-class classification to predict the number of sound events. Let $n_{max\_event}$ be the predefined maximum number of events that can occur in one time frame. For DCASE SELD dataset, $n_{max\_event}=2$. Each event has a sound class prediction, and a corresponding azimuth and elevation predictions for DOA. For each sound event, we do multi-class classification to predict sound classes, azimuths and elevations separately. We add one more \emph{background} class to the sound classes to account for segments of audio that have only background noise. During test, we first determine the number of active events $n_{active}$. After that, we pick $n_{active}$ events with the highest sound-class probabilities that are not background. The sound class and the DOA (azimuth and elevation) are the classes that have the highest probabilities. From our experiment, this output format achieves better accuracy compared to the regression format for DOA and requires fewer output neurons  to encode the DOAs ($n_{max\_event} \times (n_{azimuths} + n_{elevations})$) compared to the multi-class classification format for DOAE ($n_{azimuths} \times n_{elevations}$)~\cite{adavanne2018doaRcnn}.

\section{Experimental results and discussions}
\label{sec:majhead}


We use the DCASE 2019 SELD dataset in the FOA format~\cite{Adavanne2019_DCASE} to evaluate our approach. This dataset consists of a $400$-minute development set divided into $4$ folds and a $100$-minute evaluation set. The microphone-array signals are synthesized by convolving clean audio signals with room impulse responses recorded from five indoor locations. The maximum number of overlapping events in time is $2$ and about $50\%$ of the dataset contains $2$ overlapping events. There are $11$ sound classes, $36$ discretized azimuth angles, and $9$ discretized elevation angles, which translates to a total of $324$ possible DOAs.

\subsection{Evaluation metrics}

\begin{figure}[tb]
\centering
\includegraphics[width=0.45\textwidth]{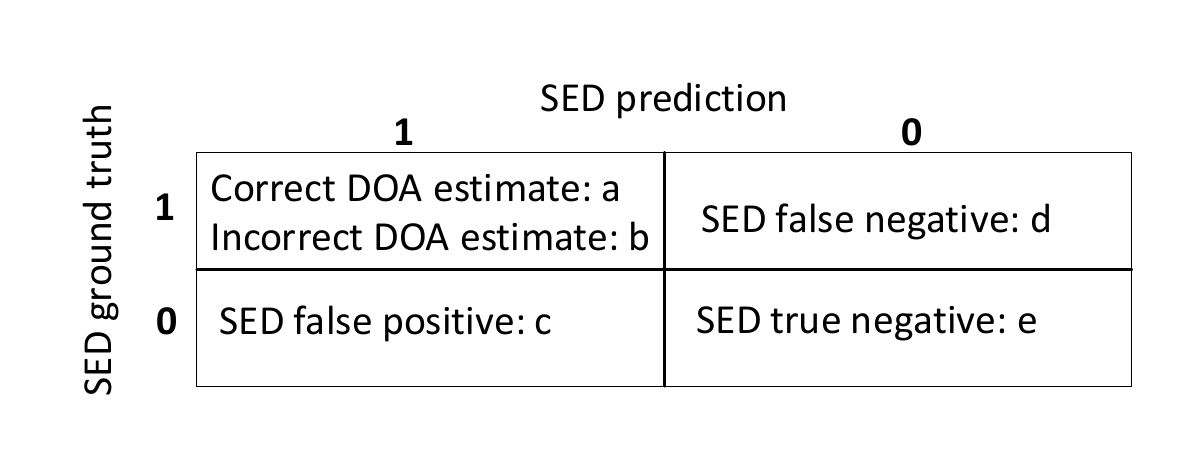}
\caption{Frame-based confusion matrix for matching SED and DOA.}
\label{fig5:matching_fscore}
\end{figure}

The SELD task is often evaluated with individual metrics for SED and DOAE~\cite{Adavanne2019seld, cao2019polyphonic, Adavanne2019_DCASE}. SED is evaluated using the segment-based F1 score and error rate~\cite{Mesaros2016_MDPI}. DOAE is evaluated using frame-based DOA error and frame recall. The segment length and the frame length are $1$ second and $0.02$ second, respectively. These evaluation metrics do not account for the matching of the DOA estimates with the corresponding sound classes. We introduce two new evaluation metrics that quantify the matching performance of a SELD system. The first evaluation metric is a frame-based matching F1 score. 
In Fig.~\ref{fig5:matching_fscore}, $a$ denotes the number of frame-based events that have correct SED predictions and correct DOA estimations within a pre-defined tolerance of $10^{\circ}$. Similarly, $b$ denotes the number of frame-based events that have correct SED predictions but incorrect DOA estimations. The frame-based matching precision and recall are $mp=a/(a+b+c)$ and $mr = a/(a+b+d)$, respectively. The frame-based matching F1 score is $mf = 2mp \times mr/(mp+mr)$. The second evaluation metric is the same-class matching accuracy that accounts only for frames that have two or more events belong to the same sound class. The same-class matching accuracy is computed as the ratio between the same-class frame-based events that have correct SED predictions and DOA estimations, and all the same-class frame-based events. A good SELD performance should have low SED error rate (ER), high SED F1 score (F1), low DOA error (DE), high DOA frame recall (FR), high matching F1 score (MF), and high same-class matching accuracy (SA). 

\subsection{Hyperparameters and training procedure}

For feature extraction, we use a sampling rate of $32$ kHz, window length of $1024$ samples, hop length of 320 samples ($10$ ms), Hann window, $1024$ FFT points, and $96$ mel filters. The sequence length of one training sample is $n_{frames} = 208$ audio frames. We use Adam optimizer to train the SED and the sequence matching networks. We train the SED network for $50$ epochs with the learning rate set to $0.001$ for the first $30$ epochs and reduced by $10\%$ for each subsequent epoch. For the single-source histogram estimation, we use 
magnitude signal-to-noise ratio of $1.5$ for the magnitude test, and a condition number of $5$ for the coherence test. We train the SMN for $60$ epochs with the learning rate set to $0.001$ for the first $40$ epochs and reduced by $10\%$ for each subsequent epoch.

\subsection{SELD baselines and the proposed two-step approach}

We denote our proposed approaches as \textbf{SMN} and \textbf{SMN-event}, which are the SMNs with the conventional and proposed SELD output format as shown in Fig.~\ref{fig4:output_format}a and Fig.~\ref{fig4:output_format}b respectively. We compare them with the following methods:

\begin{itemize}[leftmargin=*]
\item \textbf{SELDnet}: jointly trains SED and DOAE~\cite{Adavanne2019seld}, with log-mel and GCC-PHAT input features~\cite{cao2019polyphonic},
\item \textbf{Two-stage}: two-stage training strategy for SELD~\cite{cao2019polyphonic},
\item \textbf{Two-stage-aug}: is the same as the \textbf{Two-stage}, but use additional random cut-out augmentation for input features,
\item \textbf{SED-net}: same as the SED network in the \textbf{Two-stage-aug}, produce SED output sequences for SMN, 	
\item \textbf{SS-hist}: single-source histogram for DOAE to produce DOA output sequences for the SMN~\cite{tho2014robust},	
\item \textbf{Kapka-en}: the consecutive ensemble of CRNN models with heuristics rules that ranked $1$ in team category in DCASE 2019 SELD challenge ~\cite{Kapka2019seld},
\item \textbf{Two-stage-en}: the ensemble based on two-stage training that ranked $2$ in team category in DCASE 2019 SELD challenge~\cite{Cao2019DCASE}.
\end{itemize}

\subsection{SELD experimental results}

\begin{table}\small
\centering
\caption {SELD development results ($4$-fold cross validation)}  \vspace*{5pt}
\label{table1: SELD_dev}
\scalebox{0.85}{
\begin{tabular}{|c|c|c|c|c|c|c|}
\hline 
Methods & ER & F1 & DE & FR & MF & SA \\ 
\hline 
SELDnet & 0.239 & 0.866 & 11.8$^{\circ}$ & 0.836 & 0.679 & 0.201 \\ 
\hline 
Two-stage & 0.179 & 0.900 & 9.55$^{\circ}$ & 0.855 & 0.733 & 0.237 \\ 
\hline 
Two-stage-aug & 0.155 & 0.914 & 9.60$^{\circ}$ & 0.869 & 0.752 & 0.261 \\ 
\hline 
SED-net & 0.155 & 0.914 & - & - & - & - \\ 
\hline 
SS-hist & - & - & \textbf{4.44$^{\circ}$} & 0.836 & - & - \\ 
\hline 
\textbf{SMN} & \textbf{0.124} & \textbf{0.928} & 6.93$^{\circ}$ & 0.893 & \textbf{0.806} & 0.321 \\ 
\hline 
\textbf{SMN-event} & 0.128 & 0.925 & 6.27$^{\circ}$ & 0.905 & 0.774 & \textbf{0.563} \\ 
\hline 
Kapka-en & 0.14 & 0.893 & 5.7$^{\circ}$ & \textbf{0.956} & - & - \\ 
\hline 
Two-stage-en & 0.13 & \textbf{0.928} & 6.7$^{\circ}$ & 0.908 & - & - \\ 
\hline 
\end{tabular} 
}
\end{table}

\begin{table}\small
\centering
\caption {SELD evaluation results}  \vspace*{5pt}
\label{table2: SELD_eval}
\scalebox{0.85}{
\begin{tabular}{|c|c|c|c|c|c|c|}
\hline 
Methods & ER & F1 & DE & FR & MF & SA \\ 
\hline 
SELDnet & 0.212 & 0.880 & 9.75$^{\circ}$ & 0.851 & 0.750 & 0.229 \\ 
\hline 
Two-stage & 0.143 & 0.921 & 8.28$^{\circ}$ & 0.876 & 0.786 &  0.270 \\ 
\hline 
Two-stage-aug & 0.108 & 0.944 & 8.42$^{\circ}$ & 0.892 & 0.797 & 0.270 \\
\hline 
SED-net & 0.108 & 0.944 & - & - & - & - \\ 
\hline 
SS-hist & - & - & 4.28$^{\circ}$ & 0.825 & - & - \\ 
\hline 
\textbf{SMN} & \textbf{0.079} & \textbf{0.958} & 4.97$^{\circ}$ & 0.913 & \textbf{0.869} & 0.359 \\ 
\hline 
\textbf{SMN-event} & \textbf{0.079} & 0.957 & 5.50$^{\circ}$ & 0.924 & 0.840 & \textbf{0.649} \\ 
\hline 
Kapka-en & \textbf{0.08} & 0.947 & \textbf{3.7$^{\circ}$} & \textbf{0.968} & - & - \\ 
\hline 
Two-stage-en & \textbf{0.08} & 0.955 & 5.5$^{\circ}$ & 0.922 & - & - \\ 
\hline 
\end{tabular} 
}
\end{table}

Table~\ref{table1: SELD_dev} and Table~\ref{table2: SELD_eval} show the development and evaluation results, respectively. All the models except for Kapka-en and Two-stage-en are single models. First, maximizing individual performances of the SED and DOAE subtasks improves the performance of the SELD task. The random cut-out augmentation increases the SED performances of the Two-stage-aug compared to those of the Two-stage. Consequently, the SED performances of the SMNs (SMN and SMN-event) are better than those of SELDnet, Two-stage and Two-stage-aug. The single source histogram approach achieves the lowest DOA error among all the single models, which in turn improves the DOA error and frame recall of the SMNs. Using the sequence matching approach, we have the flexibility in designing SED and DOAE systems to maximize the performance of the SELD system.  


Second, the proposed SMNs outperform both the joint and two-stage training networks in all evaluation metrics. Because both the SED network and the DOAE module estimate the active periods of the detected sound events, combining the DOA and SED output sequences helps to reduce errors made by individual components and boost the overall performance. We can see that SMNs significantly reduce the SED error rate, DOA error and improve SED F1 score and DOA frame recall. Both SMN and SMN-event have similar SED and DOA performance.

Third, the proposed SMNs achieve higher matching F1 score and same-class accuracy than the joint and two-stage training models. The SMNs are able to associate the predicted sound classes and the estimated DOAs from the SED and DOAE output sequences. In addition, the SMN-event with the new proposed SELD output format has higher same-class matching accuracy than the SMN with the conventional SELD output format. The new proposed output format can scale to cases that have higher numbers of overlapping events per frame by increasing the value of $n_{max\_event}$. 

Finally, the performance of our single-model SMNs are comparable with the sophisticated and highly-tuned ensembles that rank top in the DCASE 2019 SELD challenges. Our proposed SMNs are promising for the SELD task. 




\bibliographystyle{IEEEbib}

\end{document}